\documentclass[preprint,nopacs,preprintnumbers,amsmath,amssymb,nofootinbib]{revtex4}

\usepackage{graphicx}
\usepackage{dcolumn}
\usepackage{bm}

\usepackage{bbm}

\newcommand{\bignone}{}
\newcommand{\emdash}{---}
\newcommand{\mathd}{\mathrm{d}}
\newcommand{\tmem}[1]{{\em #1\/}}
\newcommand{\tmop}[1]{\ensuremath{\operatorname{#1}}}
\newcommand{\tmstrong}[1]{\textbf{#1}}
\newcommand{\tmtextit}[1]{{\itshape{#1}}}

\begin{document}

\title{Describing viable technicolor scenarios}

\author{Johannes Hirn}
\email{johannes.hirn@yale.edu}
 \affiliation{Department of Physics, Yale University, New Haven, CT 06520}
\author{Adam Martin}
 \email{adam.martin@yale.edu}
 \affiliation{Department of Physics, Yale University, New Haven, CT 06520}
\author{Ver\'onica Sanz}
\email{vsanz@bu.edu}
\affiliation{Department of Physics, Boston University, Boston, MA
  02215\\ Department of Physics and Astronomy, York University, 4700 Keele Street, Toronto,
ON  M3J 1P3}

\begin{abstract}
  We construct an effective lagrangian for new strong interactions at the LHC,
  including as a first step the two lightest triplets of spin-1 resonances. Our
  parametrization is general enough to allow for previously unstudied
  spectrum and couplings.
   Among available frameworks to describe the spin-1
  sector, we rely on an extra-dimensional description. Our approach
  limits the number of parameters, yet is versatile enough to describe the phenomenology of
  a wide range of new
  scenarios of strong electroweak symmetry breaking.
\end{abstract}


\maketitle

\section{Introduction}

The approach of LHC turn-on has renewed interest in dynamical electroweak
symmetry breaking (DEWSB), whether in the traditional form of Technicolor
models {\cite{Weinberg:1979bn,Susskind:1978ms,Farhi:1980xs}}, or that of its
possible 5D dual description {\cite{hep-th/0012148,hep-th/0012248}} and
related moose models {\cite{hep-th/0104005,hep-th/0104179}}. Yet, only a
handful of non-supersymmetric models of electroweak symmetry breaking have so far been implemented in Monte-Carlo generators
{\cite{hep-ph/0508185,0708.2588,Adam-and-Ken}}.

To pave the way for more simulations of DEWSB, we define a flexible framework
with resonances and more generic interactions than have previously been
considered. At the same time, we strive to limit the number of parameters in
this lagrangian, to make the parameter space manageable. Dialing the parameters
then allows to describe a sampling of strong interaction models, as mSUGRA did
for the MSSM: our framework is intended to work in a similar way, applied to
DEWSB. We have already presented some phenomenological applications in
{\cite{0712.3783}}; the present paper details the inner workings of our
framework.

In practice, our approach extends attempts to model strong interactions, based
on the ideas of hidden local symmetry and mooses
{\cite{Bando:1984ej,Bando:1987br,hep-ph/0312324}} as well as 5D warped models
{\cite{hep-ph/0306259,hep-ph/0312247}} (themselves inspired by AdS/CFT
{\cite{hep-th/9711200,hep-th/9802109,hep-th/9802150}}). These approaches have
been used for QCD {\cite{hep-ph/0304182,hep-ph/0501128,hep-ph/0501218}} with
some accuracy, and thus tend to describe strong interactions which closely
follow QCD behavior up to a rescaling of $N_c$ and $\Lambda_{\tmop{QCD}}$. For
instance, they all predict the following: 1) the spectrum consists of an
alternance of states with definite parity, implying that only half of
the neutral resonances couple to $W W$ scattering, 2) the lightest
resonance has vector parity, and 3) the photon couplings to
two different particles vanish (the equivalent of $a_1 \rightarrow \pi
\gamma$ vanishing in QCD).

To explore different scenarios, we will build on a previous construction
--Holographic Technicolor (HTC) {\cite{hep-ph/0606086}}-- in which the new
strong interactions can differ from rescaled QCD. In addition to their
allowing new phenomena, deviations from QCD may help alleviate difficulties
with electroweak precision tests (EWPTs) {\cite{Peskin:1991sw}}. HTC uses 5D
language but, compared to Higgsless models {\cite{hep-ph/0308038}}, it adds
deviations from pure AdS 5D geometry in the form of effective warp factors
which differ for the various fields. As the name {\it effective warp
  factors} implies, we are only borrowing 5D language to
describe a 4D scenario. Therefore, it does not matter that simple 5D models do
not reproduce these effective warp factors by using bulk scalars
{\cite{0704.1821}}. Indeed, the same physics could be obtained in a purely 4D
context by using moose notation {\cite{Sekhar}}.

In the present paper, we use the HTC language to model the resonances and SM
gauge boson sectors. The main constraint on such interactions comes from
experimental bounds on trilinear gauge-couplings (TGCs). However, a
phenomenological study also requires modeling the fermion-resonance
interactions. In the present phenomenological description, these couplings are
simply assigned by hand: we fix the couplings between fermions and resonances to pass
current limits. (Modelling the fermions in 5D would unnecessarily increase the
number of parameters in our study. On the other hand, it may bring
interesting consequences, such as a preferential coupling of the resonances to
third generation fermions.) The present paper deals with the low-energy
lagrangian and its relation with the 5D description. Phenomenological studies
for a few benchmark points were presented previously {\cite{0712.3783}}.

We start in Section II by describing the interactions in an effective
lagrangian of spin-1 resonances. In Section III, we then discuss our HTC
framework for reducing the number of parameters as compared to a generic
effective lagrangian, and its relation to 5D modelling. In Section IV, we
detail some important properties of the model relative to the parity of
resonances. In Section V, we study the constraints that TGC bounds from LEP
impose on the parameter space of HTC. Section VI  explores the
predictions of the model for two interesting regions of parameter space.

\section{Effective lagrangian: cubic interactions}\label{sec_4D}

We consider a spectrum consisting of the SM spin-1 fields ($\gamma, W,
Z$), two triplets of resonances, as well as the SM quarks and leptons, but no
physical Higgs particle. We refer to the resonances as $(W_{1, 2}^{\pm},
Z_{1, 2})$ rather than $(\rho_T^{\pm, 0}, a_1^{\pm, 0})$ since they will turn
out to lack a definite parity, see Section \ref{sec_parity}.

In this Section, we detail the cubic couplings of dimension 4 between spin-1
fields. We restrict ourselves to  cubic vertices involving
one resonance and quartic vertices with SM fields, as is sufficient to study
the production of new heavy states at the LHC. Even at this level, we need to
make assumptions to limit the independent parameters to a manageable number.
We will use HTC to this effect in Section \ref{sec_framework}.

\subsection{Deviations from SM couplings}

The presence of a new sector affects the self-couplings of SM gauge fields,
introducing deviations from the SM in the
TGCs. In this paper, we will consider constraints from the following
TGCs {\cite{Hagiwara:1986vm}} 
\begin{eqnarray}
  \mathcal{L} & \supset & - ie \left( \left( \partial_{\left[ \mu \right.}
  W^-_{\left. \nu \right]} W^{+ \mu} A^{\nu} + \partial_{[\mu} W^+_{\left. \nu
  \right]} A^{\mu} W^{- \nu} \right) + \kappa_{\gamma} \partial_{\left[ \mu
  \right.} A_{\left. \nu \right]} W^{- \mu} W^{+ \nu} \right) \nonumber\\
  & - & ie \frac{c}{s}  \left( g_1^Z \left( \partial_{\left[ \mu \right.}
  W^-_{\left. \nu \right]} W^{+ \mu} Z^{\nu} + \partial_{[\mu} W^+_{\left. \nu
  \right]} Z^{\mu} W^{- \nu} \right) + \kappa_Z \partial_{\left[ \mu \right.}
  Z_{\left. \nu \right]} W^{- \mu} W^{+ \nu} \right),  \label{TGVs}
\end{eqnarray}
where $\kappa_{\gamma}, g_1^Z, \kappa_Z$ can differ from their SM value (equal
to 1). $c$ and $s$ in equation (\ref{TGVs}) are the cosinus and sinus of the
Weinberg angle.

From (\ref{TGVs}) above, we see that the $W W \gamma$ interaction is in
general built up of two separate Lorentz structures with independent couplings
($e$ and $e \kappa_{\gamma}$). The $W W Z$ interactions also contains two
independent couplings ($g_1^Z$ and $\kappa_Z$). Most scenarios of DEWSB impose
$\kappa_{\gamma} = 1$ and $\kappa_Z = g_1^Z$ (see Section
\ref{sec_permutation}). Current bounds on the TGCs (\ref{TGVs}) will restrict
the parameters in our description of strong interactions (Section
\ref{sec_TGCs}).

\subsection{Resonance couplings}

Introducing the spin-1 resonances as massive gauge fields generically called
$B^-, C^+, D^0$, we consider the following dimension-4 cubic
couplings
\begin{eqnarray}
  \mathcal{L} & \supset & - i \left( g_{B C D 1} \partial_{\left[ \mu \right.}
  B^-_{\left. \nu \right]} C^{+ \mu} D^{0 \nu} + g_{B C D 2} \partial_{[\mu}
  C^+_{\left. \nu \right]} D^{0 \mu} B^{- \nu} + g_{B C D 3} \partial_{\left[
  \mu \right.} D^0_{\left. \nu \right]} B^{- \mu} C^{+ \nu} \right), 
  \label{cubics}
\end{eqnarray}
where the three independent couplings $g_{B C D 1} \neq g_{B C D 2} \neq g_{B
C D 3}$ are consistent with all the low-energy symmetries. The number of free
couplings thus increases quickly as we include more resonances: we find 45 new
couplings if we limit ourselves to the $W^{\pm}, Z$ and two triplets of
resonances.

Specializing to photon couplings, i.e. $D = \gamma$, the unbroken $\mathrm{U}
\left( 1 \right)_{\tmop{em}}$ gauge invariance imposes on
(\ref{cubics}) the following relation
\begin{eqnarray}
  g_{B C \gamma} & = & 0, \hspace{1em} \tmop{if} \hspace{1em} B \neq C, 
\end{eqnarray}
while there is no constraint on the third coupling in (\ref{cubics})
\begin{eqnarray}
  g_{B C \gamma 3} & \neq & 0, \hspace{1em} \tmop{if} \hspace{1em} B \neq C . 
  \label{off-diag-photon}
\end{eqnarray}
This yields 7 cubic couplings involving the photon. The presence of a coupling
between the photon and two different particles of unequal masses, explored
recently {\cite{Adam-and-Ken,us-in-LH2007}}, produces striking signals in
collider studies {\cite{0712.3783}}.

Although we have restricted ourselves to cubic vertices, the number of
parameters is already $\mathcal{O}(50)$, far too large for a collider study.
Section \ref{sec_framework} will introduce Holographic Technicolor (HTC)
{\cite{hep-ph/0606086}}, the framework we use to reduce the number of
parameters.

\subsection{Fermion couplings}

In this paper we {\tmem{define}} the interactions of the fermions by
setting their couplings to the $W, Z$  mass eigenstates by hand to
obey the SM relations.

Usually, $S$ and $T$ are defined from two-point functions of $W$ and
$Z$, or from operators in an effective lagrangian like $W^3 B$. Here
we follow a different procedure by working directly in the mass basis
where there are no mixing operators which would
contribute to the $S$ and $T$ parameters.
Moreover, by imposing SM couplings between fermions and  $W$, $Z$ and
$\gamma$, we ensure that the amplitudes extracted experimentally
satisfy $S=T=0$. This can be read off from the expression for the neutral current amplitude \cite{Chivukula:2004pk}
\begin{eqnarray}
{\cal M}_{NC} &=& e^2 \frac{{\cal Q Q'}}{Q^2}+ \frac{(I_3-s^2 {\cal Q})
  (I'_3-s^2 {\cal Q'})}{\frac{1}{4 \sqrt{2} G_F}-s^2 c^2 M_Z^2
  \frac{T}{4 \pi}+\left(  \frac{s^2 c^2}{e^2}-\frac{S}{16 \pi} \right)
  Q^2} + \text{non-oblique contributions}.
\end{eqnarray}

Our phenomenological study is not intended to
present a UV completion that would resolve the clashes between DEWSB and
oblique corrections. Rather, it tries to present the possible phenomenological
consequences of a scenario that would pass the oblique and TGC constraints
(Section \ref{sec_TGCs}).

We set the couplings of fermions to the resonances to be compatible with
experimental bounds from LEP and Tevatron
{\cite{hep-ph/0106251,hep-ex/0503048,hep-ex/0611022,hep-ex/0702027,0707.2524,CDFnote8452,CDFnote9150}}.

\section{5D with a twist}\label{sec_framework}

To keep the study manageable in the spin-1 sector, we use 5D techniques,
trading the plethora of resonance couplings in (\ref{cubics}) for a few
extra-dimensional parameters.

The usual reason for using 5D models to describe 4D strongly interacting
theories is the AdS/CFT correspondence
{\cite{hep-th/9711200,hep-th/9802109,hep-th/9802150}}. Although rigorous
derivations of this duality have only been obtained for very specific cases,
we do not need an exact equivalence in order to study LHC phenomenology. Indeed,
provided the essential properties of the 5D model are the same as the strong
interaction scenario, we can use the 5D description as a physical guide and
organizing scheme. The 5D description also allows for the introduction of
deviations from rescaled QCD, using for instance Holographic Technicolor
(Section \ref{sec_HTC}).

\subsection{5D Basics}

We quickly review the extra-dimensional properties and language we need. The
geometry of the extra-dimension is described by the warp factor $w (z)$, as in
the line element
\begin{eqnarray}
  \mathd s^2 & = & w (z)^2  \left( \eta_{\mu \nu} \mathd x^{\mu} \mathd
  x^{\nu} - \mathd z^2 \right) . 
\end{eqnarray}
The $z$-coordinate is finite, extending from $l_0$ (UV brane) to $l_1$ (IR
brane). A gauge field propagating in the 5D space-time, $A_M (x^{\mu},
z)$possesses five indices $M = (\mu, 5)$, and can be decomposed as an infinite
sum of 4D excitations
\begin{eqnarray}
  A_M (x, z) \text{$_{}$} & = & \sum^{\infty}_n a^{(n)}_M (x^{\mu})
  \varphi^{(n)}_M (z) .  \label{KKred}
\end{eqnarray}
$\varphi_M^{(n)} (z)$ is the wavefunction, or profile, of the 4D field
$a_M^{(n)} (x)$ along the extra-dimension $z$ (in the {\tmem{bulk}}). This
Fourier decomposition is called in this context {\tmem{Kaluza-Klein}} (KK)
decomposition, and the infinite tower of KK excitations, the KK-tower.

The wavefunctions $\varphi_M^{(n)} (z)$ are obtained by solving the equation
of motion of the field $A_M$ in the background given by $w (z)$. The
wavefunctions also depend on the boundary conditions (BCs) imposed at $l_0$
and $l_1$. One can choose BCs to partly or completely break the 5D gauge
symmetries at low energies. Specifically, if the BCs do not allow massless
(i.e. flat) modes in the spectrum, then there is no remaining 4D gauge
invariance.

Once the wavefunctions are known, the interactions between 4D fields can be
derived from overlap integrals. For example, the coupling of the $Z$ and $W$
bosons to a resonance $W_i$ is the integral of their wavefunctions along the
$z$-coordinate, \
\begin{eqnarray}
  g_{ZW \rho} & \propto & \int_{l_0}^{l_1} \frac{\mathd z}{g^2_5}  w(z)
  \varphi_Z \varphi_W \varphi_{W_i} \bignone \ldots .  \label{tri0}
\end{eqnarray}
Such couplings can be computed easily, yielding from the
lagrangian (\ref{Linit}) the value of all the cubic couplings of SM gauge
fields (\ref{TGVs}) and resonances (\ref{cubics}).

For a given choice of gauge group, BCs and geometry, a 5D model uses only a
few pararmeters to describe a complex scenario of many particles. In the
simplest version (AdS), these parameters are: the length
$l_1$, the dimensionless gauge coupling $l_0 / g_5^2$ and the form of
the geometry $w(z)=\frac{l_0}{z}$. In the following we introduce {\it two}
more parameters in the functional shape of the warp factors. These two
new parameters are an essential ingredient to achieve a
departure from QCD-like physics (Section \ref{nonQCD}) while
maintaining 5D relations such as (\ref{tri0}).

\subsection{The lagrangian of Holographic Technicolor}\label{sec_HTC}

To model DEWSB, we place $\tmop{SU} (2)_L \otimes \tmop{SU} (2)_R$ gauge
fields in the 5D bulk. The lightest KK excitations of these 5D gauge fields
correspond to the SM elecroweak gauge bosons, while the higher KK excitations
of the same 5D fields will be interpreted as resonances. HTC corresponds to
the following choice of bulk action
\begin{eqnarray}
  S & = - \frac{1}{4 g^2_5} & \int \mathd^4 x \mathd z \frac{w_V \left( z
  \right) + w_A \left( z \right)}{2} \left( F_{L, M N}^a F_L^{a, M N} + F_{R, M N}^a
  F^{a, M N}_R \right) \nonumber\\
&+& \left( w_V \left( z \right) - w_A \left( z \right) \right) F_{L, M N}^a
  F_{R^{}}^{a, M N},  \label{Linit}
\end{eqnarray}
where $a$ labels the $\tmop{SU} (2)$ generator. We chose to write a bulk
lagrangian invariant under $L \leftrightarrow R$, i.e. parity. In that case,
it is convenient to work in terms of the vector and axial combinations of
gauge fields $V, A = \left( R \pm L \right) / \sqrt{2}$, to get
\begin{eqnarray}
  S & = - \frac{1}{4 g^2_5} & \int \mathd^4 x \mathd z \left( w_V (z) F_{V,
  M N}^a F_V^{a, M N} + \omega_A (z) F_{A, M N}^a F^{a, M N}_A \right), \nonumber
\end{eqnarray}
where
\begin{eqnarray}
  F_{V, M N} & = & \partial_M V_N - \partial_N V_M - \frac{i}{\sqrt{2}} 
  \left( [V_{M,} V_N] + \left[ A_M, A_N \right] \right), \\
  F_{A, M N} & = & \partial_M A_N - \partial_N A_M - \frac{i}{\sqrt{2}} 
  \left( [V_M, A_N] + \left[ A_M, V_N \right] \right) . 
\end{eqnarray}

We use Higgsless BCs in the IR
\begin{eqnarray}
  A^a |_{\tmop{IR}} \hspace{1em} = \hspace{1em} 0, &  & \partial_z V^a
  |_{\tmop{IR}} \hspace{1em} = \hspace{1em} 0 .  \label{BCsIR}
\end{eqnarray}
The UV BCs appropriate for EWSB break parity, which will have consequences
later on (Section \ref{sec_parity})
\begin{eqnarray}
  \partial_z (V^a - A^a) |_{\tmop{UV}} &=& 0, \\
  V^{1, 2} + A^{1, 2} |_{\tmop{UV}} &=& 0,\\
 \Delta \partial_z (V^3 + A^3) |_{\tmop{UV}} &=& g_5^2 \Box_4  \left( V^3 + A^3 \right) |_{\tmop{UV}}, 
  \label{BCsUV}
\end{eqnarray}
The third BC in (\ref{BCsUV}) is achieved by adding a
brane-localized $\mathrm{U} \left( 1 \right)$ kinetic term
{\cite{hep-ph/0510275}}
\begin{eqnarray}
  - \frac{1}{4 \Delta}  \left( \partial_{\left[ \mu \right.} R^3_{\left. \nu
  \right]} \partial^{\left[ \mu \right.} R^{\left. \nu \right] 3} \right), & 
  & 
\end{eqnarray}
and requiring that the variation of the action vanish under any variation of
$R^3$. This allows for a $\mathrm{U} (1)_{B - L}$ brane
field.{\footnote{A bulk $\mathrm{U}(1)_{B-L}$ would introduce more
    neutral resonances, as in Higgsless models. In either case, an
    extra parameter enters the lagrangian.}}

The above combination of BCs ensures that the only surviving symmetry at low
energies is $U (1)_{\tmop{em}}$. Other than the photon massless mode, the
spectrum contains the massive $Z \tmop{and} W^{\pm}$ and an infinite tower of
heavier resonances $(W_i^{\pm}, Z_i)$. For given warp factors $w_A \left( z
\right), w_V \left( z \right)$, $l_1$ sets the mass of the lightest resonance
in the KK tower with respect to $M_W$. We will be interested in cases where
the lightest resonance is lighter than $1 \tmop{TeV}$, and we will truncate
the KK tower after the lightest two triplets. One consequence of adding a
sizeable $F_L F_R$ term in (\ref{Linit}) is that the two first triplets can
have comparable masses, and therefore must both be kept in the spectrum.

\subsection{Breaking patterns}\label{nonQCD}

As mentioned above, 5D models in AdS, in which there is only one warp factor
with a fixed expression $w_A (z) = w_V (z) = \frac{l_0}{z}$, exhibit features
similar to a rescaled version of QCD (alternating spectrum with selection
rules for couplings). Allowing two different warp factors which deviate from AdS
in the infrared, as in HTC (\ref{Linit}), lifts these restrictions.

For computations, we need to pick an explicit expression form for the warp
factors: we choose the positive-definite
\begin{eqnarray}
  w_{A, V} (z) & = & \frac{l_0}{z} e^{\frac{o_{A, V}}{2}  \left( \frac{z -
  l_0}{l_1} \right)^4} .  \label{wAV}
\end{eqnarray}
We use the name {\tmem{effective warp factors}} because $w_{A, V}$ do not
correspond to an actual geometry in 5D {\cite{0704.1821}}. Rather, $w_{A, V}$
parameterize different scenarios of technivectors: varying $o_{V, A}$ amounts
to changing the masses and couplings of the vector and axial resonances. The
exact power of $z$ in (\ref{wAV}) is not crucial: a modification of the power
could be partly absorbed in a modification of $o_V, o_A$. What matters is that
the effective metrics deviate from AdS in the IR, and thus modify the
wave-functions and overlap integrals (\ref{tri0}) from which the couplings are
obtained.

Beyond the breaking by IR BCs, familiar from Higgsless theories
{\cite{hep-ph/0305237}}), the different effective metrics felt by the $V$ and
$A$ fields introduce symmetry breaking in the bulk ($z$-dependence).
Brane-breaking corresponds to the choice of BCs in (\ref{BCsIR}), while
bulk-breaking is introduced on top of (\ref{BCsIR}) as different effective
metrics in (\ref{Linit})
\begin{eqnarray*}
  \text{brane-breaking} & \Longrightarrow & w_A (z) = w_V (z),\\
  \text{bulk-breaking} & \Longrightarrow & w_A (z) \neq w_V (z) .
\end{eqnarray*}
In the case of brane-breaking, the only distinction between the broken (axial)
and unbroken (vector) generators comes from the IR BCs. In a ``dual''
interpretation, where the fifth space coordinate is inversely related with an
energy scale, this corresponds to the symmetry breaking occuring suddenly at
the resonance scale {\cite{hep-th/0012248}}. The localization of the breaking
at a point in the extra-dimension suppresses effects at any scale above that.
Such hard-wall breaking is the crudest 5D model of spontaneous symmetry
breaking.

If the symmetry breaking is turned on progressively along the 5th dimension
rather than at the IR brane alone, we have bulk breaking. The strength of the
breaking, and therefore the difference between the properties of vector and
axial states, is governed by the $z$-dependence of the breaking term. The
standard way to accomplish bulk breaking is to add a scalar bulk field, and to
let it obtain a vacuum expectation value. However, the KK decomposition of the
bulk scalar would introduce 4D scalar resonances, which we want to avoid for
simplicity in the present paper. Rather, in the HTC lagrangian (\ref{Linit}),
we introduced a position dependent kinetic term mixing $L$ and $R$ gauge
fields, which is invariant under the vector gauge symmetry and parity.
Obviously, many more terms besides $F_L F_R$ could be added to the lagrangian
and may lead to different phenomenology, but $F_L F_R$ is the operator with
lowest dimension and number of derivatives and no new fields.

From the effective field theory point of view, what matters is that
bulk-breaking allows a more general spectrum and structure for the cubic
interactions of (\ref{cubics}), see Section \ref{sec_permutation}.

\subsection{Dialing the effective warp factors}

Our parameterization of the spin-1 sector is economical. Three combinations of
parameters are set by imposing the physical value of $M_W, M_Z,
\alpha_{\tmop{em}}$. The remaining parameters can be chosen as: the size $l_1$
of the extra dimension (closely related to the mass of the lightest resonance,
$M_{W_1}$) and the two parameters $o_V$ and $o_A$ describing the functional
shape of the effective background felt by the spin-1 fields.

For the fermion sector, we choose the couplings of the fermions to the $W, Z,
\gamma$ to follow the SM relations. As to the couplings of fermions to
resonances, they do not influence the results of this paper, as long as these
are suppressed by a factor compared to the couplings to SM gauge fields.{\footnote{In the phenomenological study {\cite{0712.3783}}, we set the
couplings between any fermions and a resonance $W_{1, 2}$ or $Z_{1, 2}$ to be
equal to the coupling between the same fermions and the $W$ or $Z$ respectively, rescaled
by a common factor $\kappa$, independent of the fermion or the resonance
multiplet. For the specific points studied in {\cite{0712.3783}} with 500 and
600 GeV resonances, we chose $\kappa = 1 / 20$ and $\kappa = 1 / 10$ in order
for the resonances to have avoided detection at TeVatron
{\cite{hep-ph/0106251,hep-ex/0503048,hep-ex/0611022,hep-ex/0702027,0707.2524,CDFnote8452,CDFnote9150}}.\label{footnote-about-kappa}}}

\begin{figure}[hbt]
  \includegraphics[width=10cm]{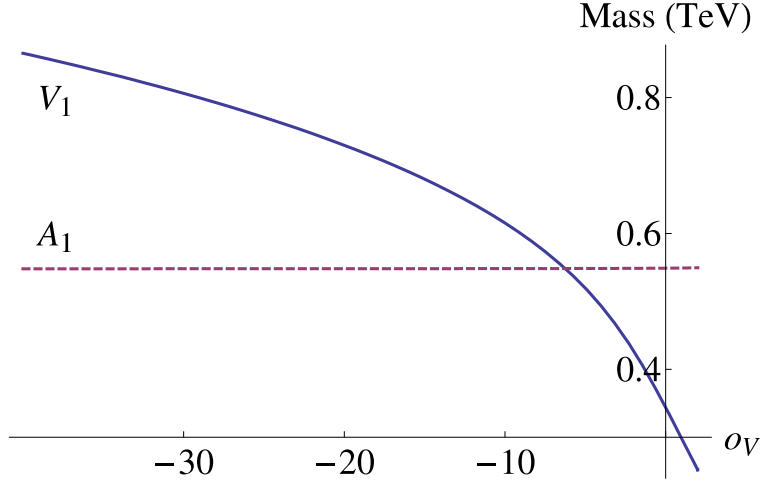}
  \caption{\label{fig_level-crossing}Mass of the lightest vector ($V_1$) and
  lightest axial ($A_1$) resonances as a function of $o_V$, for $o_A = 0$,
  $l_1 = 7 \tmop{TeV}^{- 1}$. This plot uses Dirichlet BCs in the UV.}
\end{figure}

Let us point out the basic effects of varying one of the parameters $o_V$. A
negative $o_V$ in the effective warp factor (\ref{wAV}) acts as an IR cutoff,
effectively shortening the space in which $V$ fields live, but leaving axial
masses untouched. Figure \ref{fig_level-crossing} shows this explicitely in a
simpler scenario with UV Dirichlet BCs for $V$ and $A$ fields ($V = A = 0$ at
the UV brane). With such BCs, the first axial resonance would be lighter than
the vector one for $o_V < - 5$. For applications to EWSB, the situation is
more complicated due to the different BCs, see Section \ref{sec_parity}.

\section{Parity of resonances}\label{sec_parity}

The HTC lagrangian (\ref{Linit}) is invariant under parity ($L \leftrightarrow
R$). One would thus expect the spin-1 resonances to have definite parity. In
the 5D language, the eigenfunctions would then split into two distinct
sectors: $V$ or $A$ wavefunctions, not admixtures of them. However, the
coupling to electroweak $\tmop{SU} \left( 2 \right)_L \otimes \mathrm{U}
\left( 1 \right)_Y$ interactions breaks parity. In the 5D language, this
effect comes from the UV BCs which mix the vector and axial sectors
(\ref{BCsUV}).

\begin{figure}[hbt]
  \includegraphics[width=10cm]{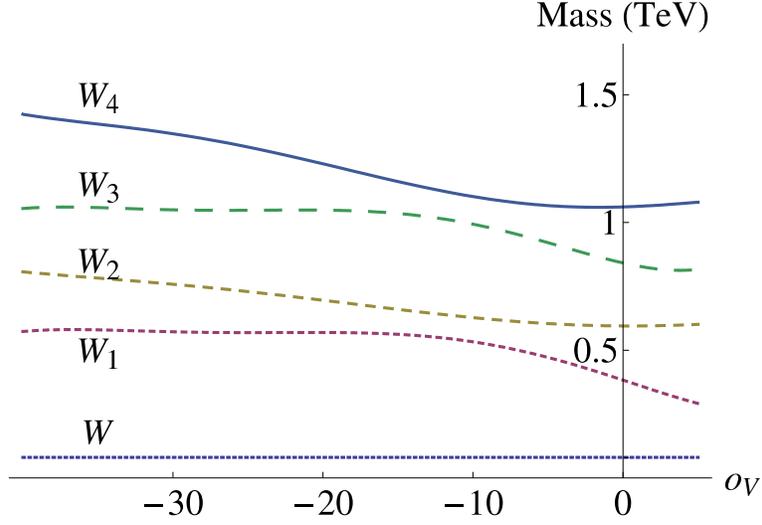}
  \caption{\label{fig_level-repulsion}Masses of the four lightest charged
  resonances in TeV, as a function of $o_V$, for $o_A = 0$, and fixing $l_1 = 7 \tmop{TeV}^{- 1}$ with
  $M_W$ at its physical value.}
\end{figure}

The mixing effects depicted in Figure \ref{fig_level-repulsion} become
especially important for nearly-degenerate resonances, the region we want to
look at. This mixing is not an artifact of our framework and will hold for any
model of nearly-degenerate resonances coupled to the electroweak sector. As
can be seen in Figure \ref{fig_level-repulsion}, the level repulsion also
affects the higher KK modes.

\begin{figure}[hbt]
  \includegraphics[width=10cm]{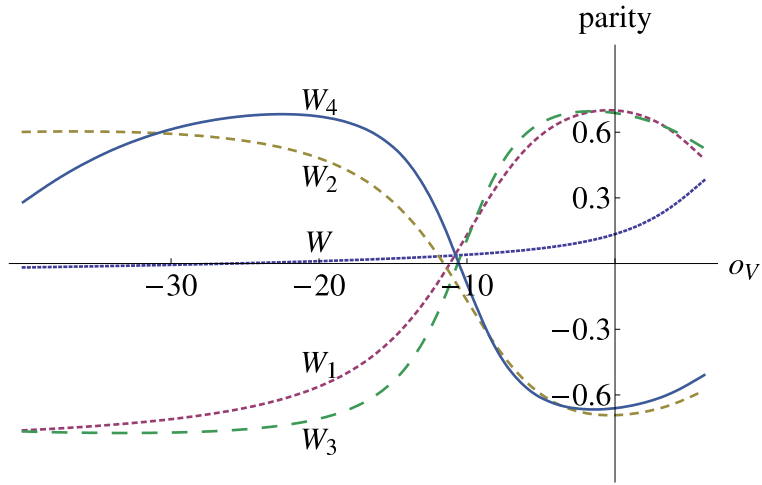}
  \caption{\label{fig_parity}Parity, as defined in (\ref{parity}), of the
  various charged modes as a function of $o_V$, for $o_A = 0$ and $M_{W_1} =
  500 \tmop{GeV}$, and $l_1 = 7 \tmop{TeV}^{- 1}$.}
\end{figure}

Due to the mixing, each eigenstate is a linear combination of a vector and an
axial wavefunction, without definite parity. The eigenfunctions are thus
two-component objects $| \Psi \rangle = |V_{\Psi}, A_{\Psi} \rangle$, where
$V_{\Psi}$ and $A_{\Psi}$ are the vector and axial component. To follow what
happens as we vary the effective warp factors, we define the continuous parity
of eigenstate $\Psi$ by
\begin{eqnarray}
  \tmop{parity} (\Psi) & = & \frac{\int \mathd z (w_V \bignone V_{\Psi}^2 -
  w_A A_{\Psi}^2)}{\int \mathd z (w_V \bignone V_{\Psi}^2 + w_A A_{\Psi}^2)}, 
  \label{parity}
\end{eqnarray}
and plot the parity of the states as we vary $o_V$ in Figure
\ref{fig_parity}.

Figure \ref{fig_parity} shows that the lightest resonance goes adiabatically
from a mostly-vector state for $o_V \approx 0$, to a mostly axial one for
large negative $o_V$, and vice-versa for the second level. Figure
\ref{fig_parity} also shows that the $W$, whose UV BC (\ref{BCsUV}) imposes it
to be predominantly $V - A$ in the UV, contains nearly-equal admixtures of $V$
and $A$ for most of the parameter space. The resonances change parity near
$o_V \simeq - 10$.

The fact that $V$ and $A$ mix to yield mass eigenstates without definite
parity will have important consequences later on in Section
\ref{sec_permutation} for the scenarios we consider. On the other hand, the
mixing is not relevant for Higgsless models, since it decreases with the mass
separation between states.

\section{Bounds from TGCs}\label{sec_TGCs}

In the present Section, we examine the limits set by TGCs on resonance masses,
and point to regions in the remaining two-dimensional parameter space $(o_V,
o_A)$ where $M_{W_1} \leqslant 500 \tmop{GeV}$ is allowed.

To avoid numerical difficulties encoutered when studying the whole parameter
space, we restrict ourselves to two curves within the $(o_A, o_V)$ plane,
along which we illustrate the constraints from TGCs in this Section. Our
choices are: line A ($o_A = 0$) and line B ($\alpha_1 = 0$), where
$\alpha_1$ is the Longhitano coefficient of the $W^3 B$ term (See Appendix
\ref{app} for details).

\begin{figure}[hbt]
  \includegraphics[width=10cm]{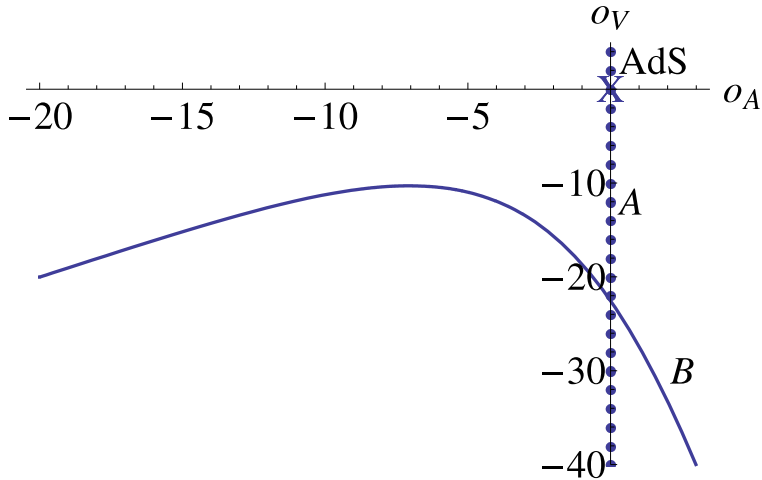}
  \caption{\label{fig_alpha1_0}The $(o_A, o_V)$ plane, with lines A and B
  (dotted and continuous, respectively) studied in the text. The approximation
  $l_0 \rightarrow 0$ has been used to determine line B. The origin of the
  axes corresponds to the pure AdS Higgsless model.}
\end{figure}

The two curves (line A and line B) are depicted in the $(o_A, o_V)$ space in
Figure \ref{fig_alpha1_0}, where the origin of the plot represents AdS space.
The curve aso shows how the physical predictions depend not only on the
difference $o_A - o_V$, but on both effective warp factors.

Line B imposes a relation between $o_A$ and $o_V$, and we will plot the
various couplings as a function of $o_A$ along that curve.

\subsection{Line A: $o_A = 0$}\label{sec_vertical}

\begin{figure}[hbt]
  \includegraphics[width=16cm]{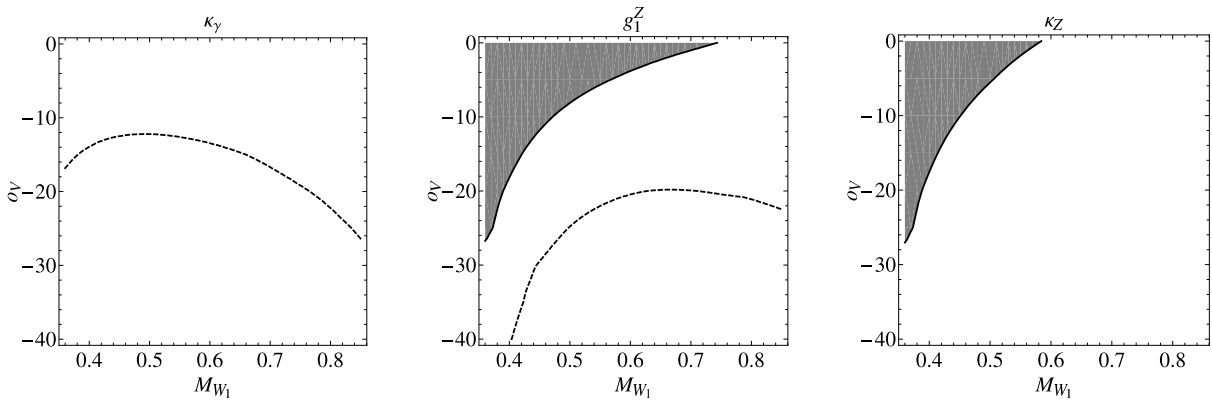}
  \caption{\label{fig_TGCs-constraints}$2 \sigma$ bounds imposed by
  $\kappa_{\gamma}, g_1^Z, \kappa_Z$ in the space of $M_{W_1}$ (horizontal
  axis) and $o_V$ (vertical axis), along line A ($o_A = 0$). Darker areas are
  excluded, and the central value is depicted as a dashed line.}
\end{figure}

The TGCs are the main constraint on HTC, since oblique corrections $\left( S,
T \right)$ vanish by our choice of fermion couplings to gauge fields. In
Figure \ref{fig_TGCs-constraints}, we depict the $2 \sigma$ constraints on
$M_{W_1}$ imposed by the bounds on each of the TGCs. Figure
\ref{fig_TGCs-constraints} shows that resonances as light as 500 GeV are
allowed by the TGC constraints, but only for $o_V \lesssim - 10$. On the other
hand, we see that a model in a pure AdS background ($o_A = o_V = 0$) is
incompatible with the TGC bounds at $2 \sigma$ unless $M_{W_1}$ is raised
above 700 GeV.

\subsection{Line B : $\alpha_1 = 0$}\label{sec_alpha1}

We study the line for which the Longhitano coefficient $\alpha_1$ vanishes.
Our motivation for studying this line is twofold. First, a vanishing
$\alpha_1$ limits the deviations of the TGCs from the SM, see
(\ref{kappagamma}-\ref{kappaZ}) in Appendix \ref{app}. This is illustrated in
Figure \ref{fig_TGCs-alpha1}, where the TGCs are plotted along the line of
$\alpha_1 = 0$, for $M_{W_1} = 500 \tmop{GeV}$. Additionally, since $\alpha_1$
describes the mixing between $W^3$ and the hypercharge gauge field $B$, $\alpha_1
= 0$ means that no additional contributions are required to cancel the $S$
parameter.

\begin{figure}[hbt]
  \includegraphics[width=17cm]{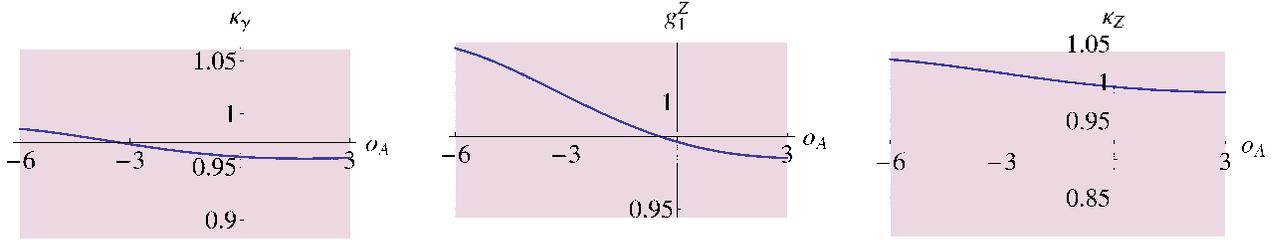}
  \caption{\label{fig_TGCs-alpha1}TGCs along line B, for $M_{W_1} = 500
  \tmop{GeV}$, with all other parameters at their physical value. The
  horizontal axis depicts the central value measured at LEP, while the shaded
  bands depict the $2 \sigma$ errors.}
\end{figure}

\section{Spectrum and couplings}\label{sec_exploring}

In this section, we set the mass of the lightest resonance to $M_{W_1} = 500 \tmop{GeV}$. We
then extract predictions for resonance couplings along the two lines A and B
described in Section \ref{sec_TGCs}. This particular choice is arbitrary, but
exemplifies the phenomenology that can be obtained in HTC.

In practice, we will truncate the KK tower, explicitly keeping in the
lagrangian only the first two multiplets of resonances: with the lightest
resonance at 500 GeV, the third multiplet would usually come in above one TeV
(see Figure \ref{fig_level-repulsion}), still interesting for the LHC purposes
but outside the scope of the present paper.

\subsection{Spectrum}\label{sec_spectrum}

The splitting between multiplets determines the decays of the heavy resonance.
A splitting larger than $\sim 100$ GeV allows for the  $W_2$ to decay into $W_1$.

\begin{figure}[hbt]
  \center{\includegraphics[width=7cm]{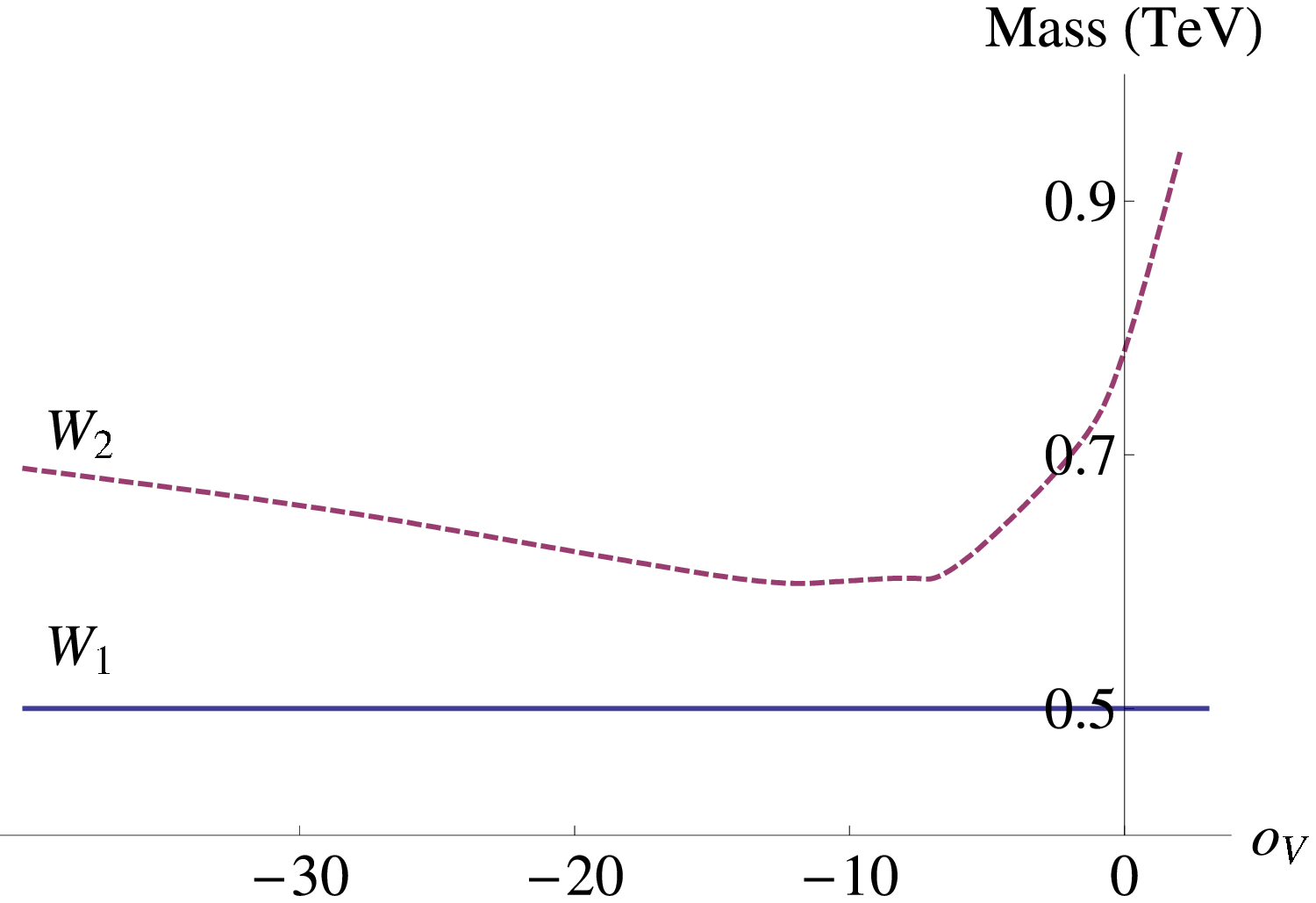}{\hspace{2em}}\includegraphics[width=7cm]{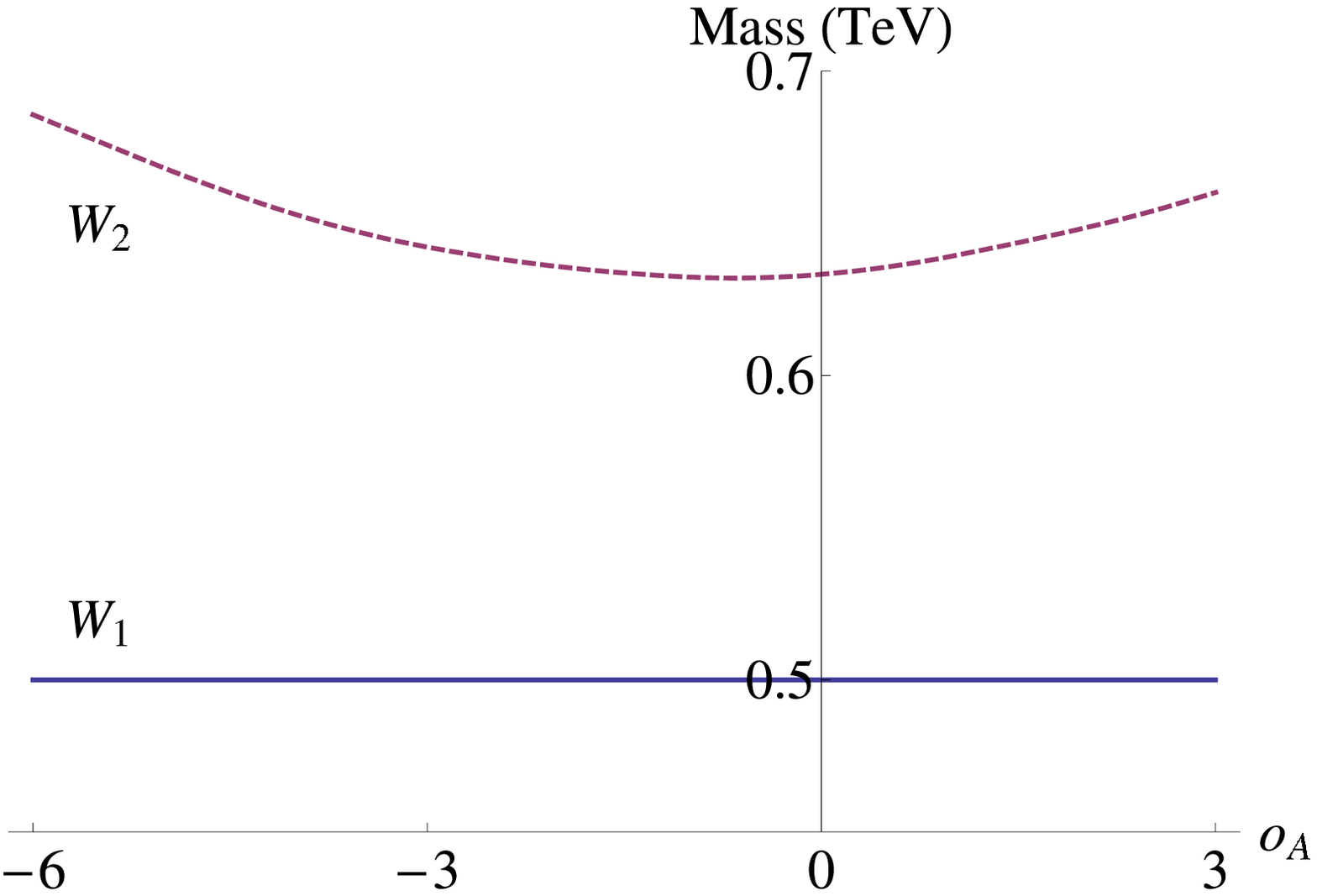}}
  \caption{\label{fig_spectrum}Variation of $M_{W_2}$ along the two curves in
  parameter space, when $M_{W_1}$ is set to $500 \tmop{GeV}$. Other parameters
  are set to reproduce the physical values of $\alpha, M_W, M_Z$.}
\end{figure}

Figure \ref{fig_spectrum} depicts the mass of $W_2$ when $M_{W_1}$ is set to
$500 \tmop{GeV}$. The left plot for line A corresponds to the one of Figure
\ref{fig_level-repulsion} above, but rather than keeping $l_1$ fixed, it is
rescaled to maintain $M_{W_1} = 500 \tmop{GeV}$. Along this line (line A), the
splitting between resonances decreases from the AdS ($o_V = o_A = 0$) as we
dial a non-zero value of $o_V$. Along line B ($\alpha_1 = 0$), the separation
between the two lightest resonances remains about 150 GeV.

\begin{figure}[hbt]
  \center{\includegraphics[width=7cm]{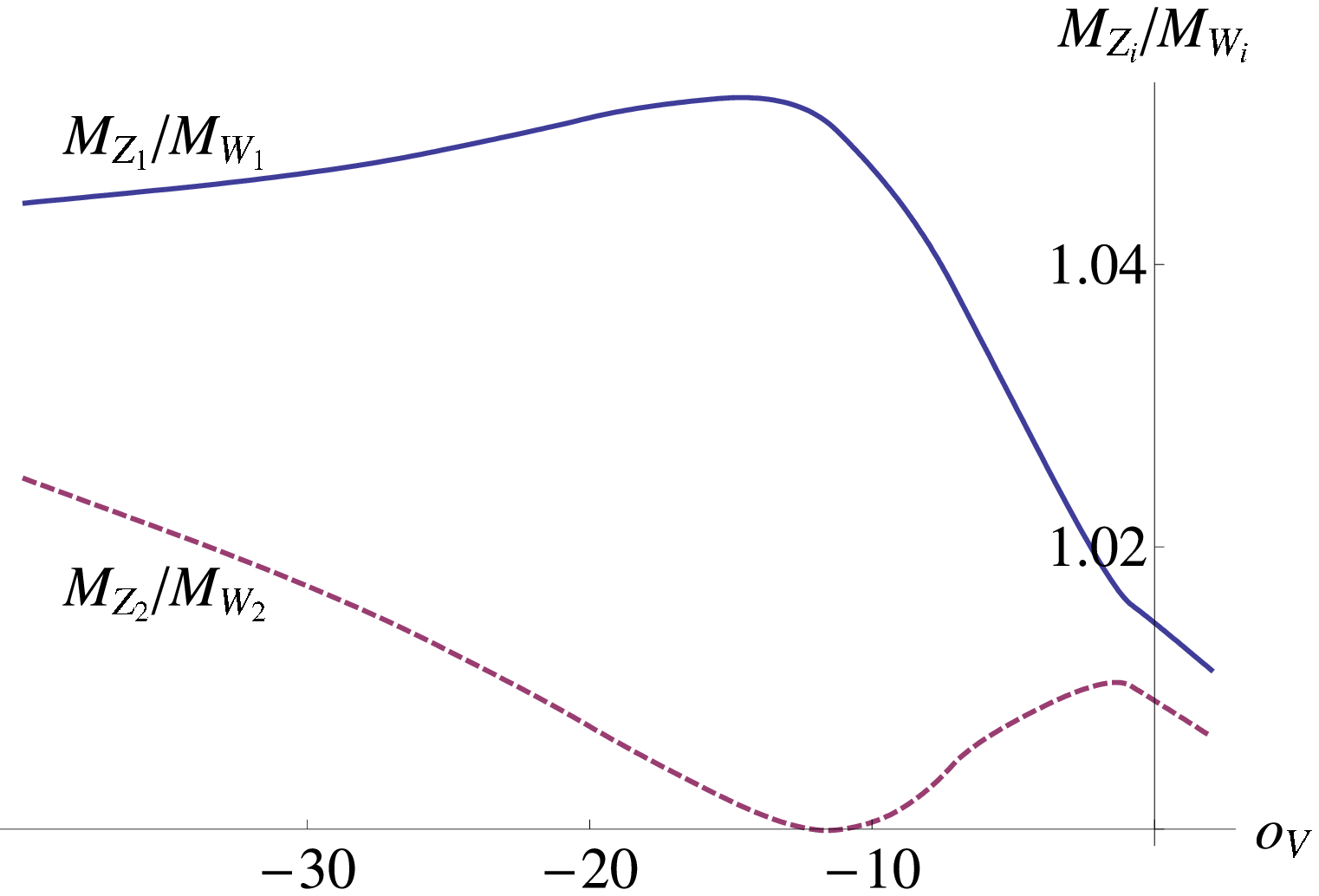}{\hspace{2em}}\includegraphics[width=7cm]{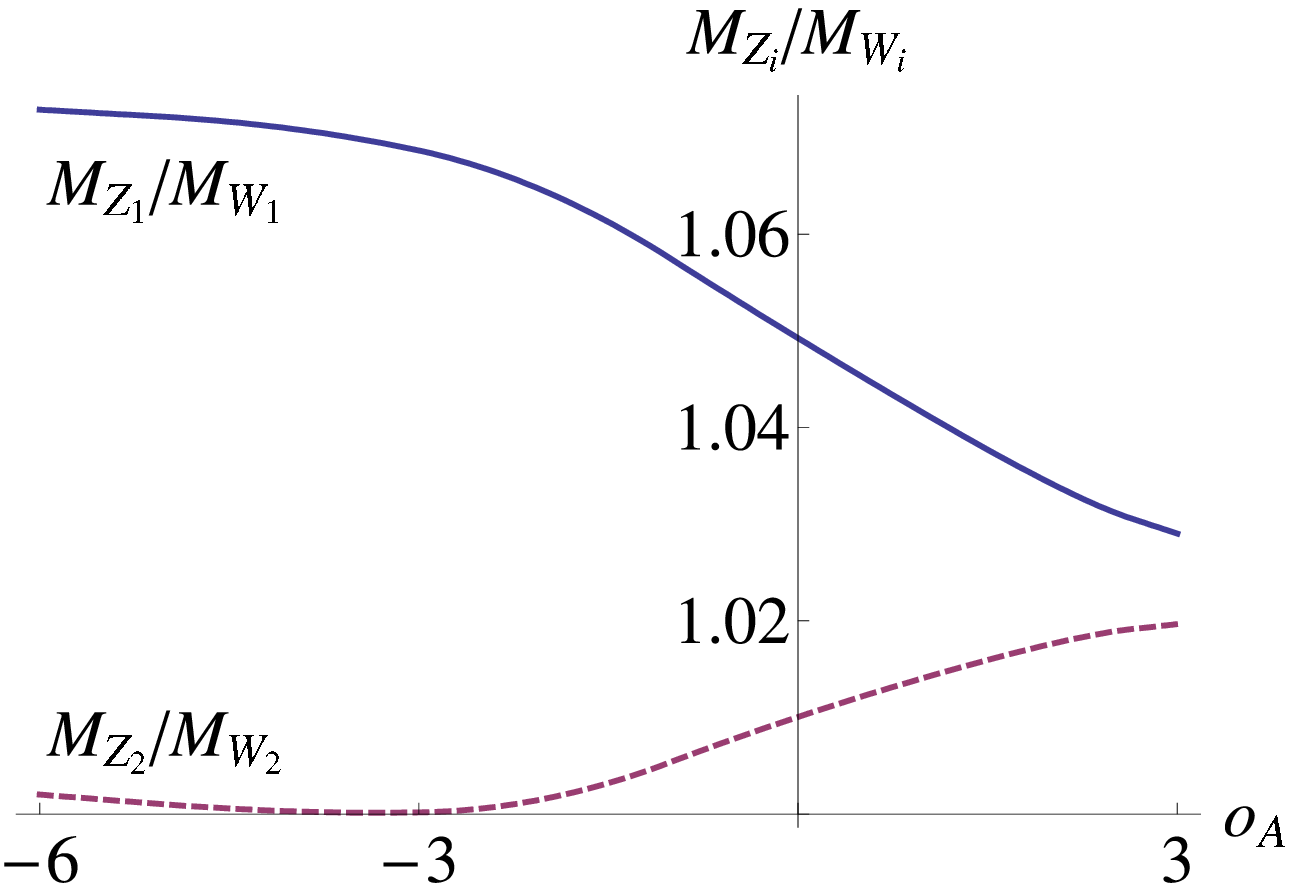}}
  \caption{\label{fig_isospin}Ratios of masses $M_{Z_1} / M_{W_1}$ and
  $M_{Z_2} / M_{W_2}$ along the the two curves in parameter space, for
  $M_{W_1} = 500 \tmop{GeV}$. Other parameters are set in order to reproduce
  the physical values of $\alpha, M_W, M_Z$.}
\end{figure}

Along the same two lines, we can examine the isospin splitting within the
lightest two triplets of resonances, as plotted in Figure \ref{fig_isospin}.
This mass splitting affects the $\rho$ parameter, which will remain close to
unity as long as the couplings of fermions to resonances are small
enough.{\footnote{We do not compute $\rho$ here, as it depends on the value of
these couplings (in particular, it depends on $\kappa$ in the example
mentioned in footnote \ref{footnote-about-kappa}). Since we have already
assigned SM couplings between the fermions and the $W, Z$, the $T$ parameter
vanishes.}} The neutral resonances are always heavier than the charged ones,
due to the UV BC (\ref{BCsUV}), which lifted the $Z$ with respect to the
$W$.

\subsection{Coupling of resonances to $W W$: new
contributions}\label{sec_pi-pi}

In this section, we discuss how a selection rule that was valid in previous
models does not apply to HTC. In the limit where the SM gauge couplings
vanish{\tmstrong{}}, the system recovers a $L \leftrightarrow R$ symmetry under
which all resonances are either even or odd under parity. In this
parity-limit, only the vector resonances would couple to two $W$'s
\begin{equation}
  g_{VW_L W_L } \neq 0, g_{A W_L W_L} = 0,
\end{equation}
and thus only the vector resonances help unitarizing $W_L W_L$ scattering.

As seen in Section \ref{sec_parity}, once the SM gauge interactions are turned
on, the $L R$ parity is violated: the axial or vector labels become
meaningless as both towers of resonances acquire both components. Given that
all resonances have a vector component, they all couple to the $W$ and $Z$.
This means that both sets of resonances participate in unitarizing
longitudinal gauge boson scattering.

\begin{figure}[hbt]
  \center{\includegraphics[width=7cm]{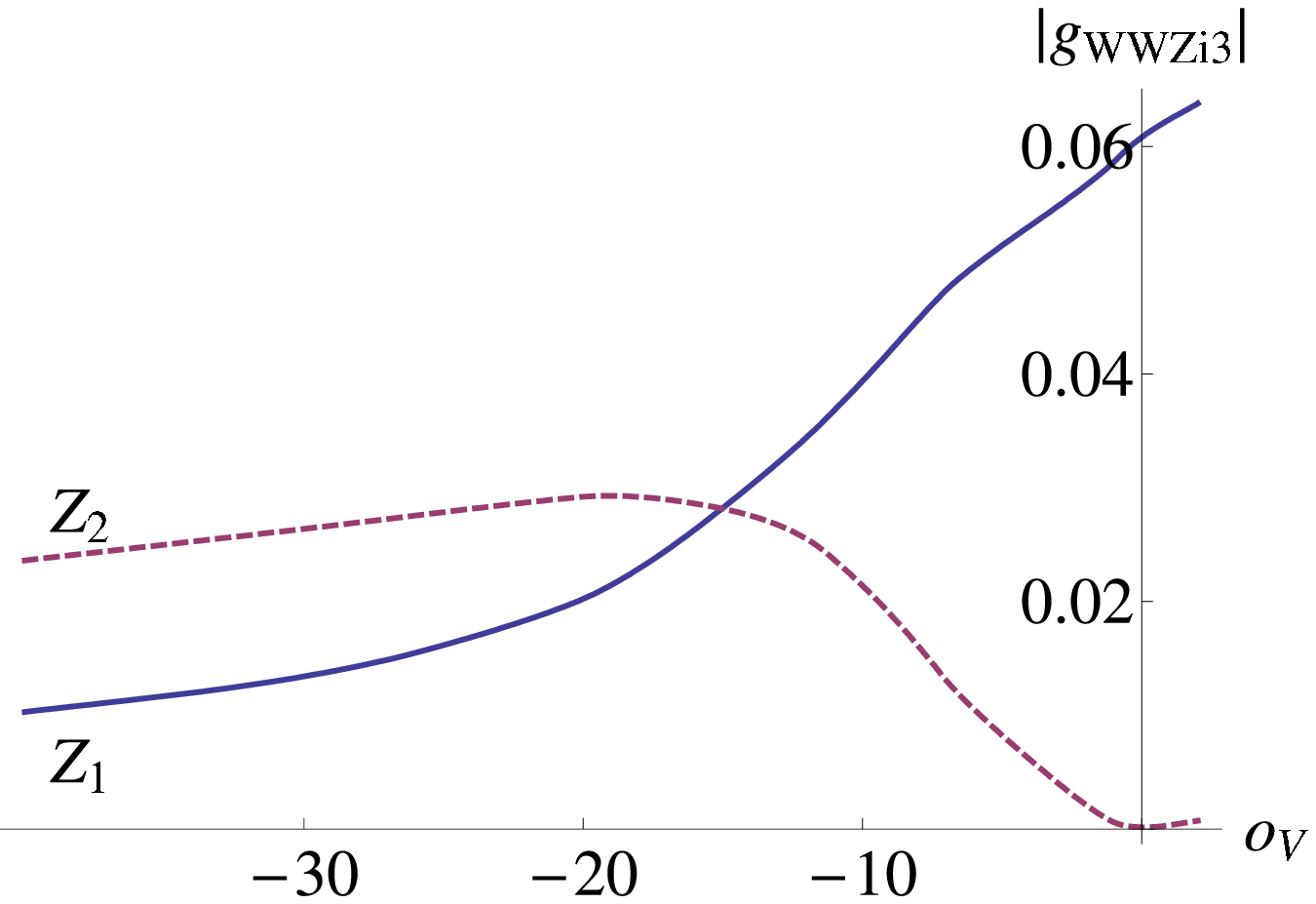}{\hspace{2em}}\includegraphics[width=7cm]{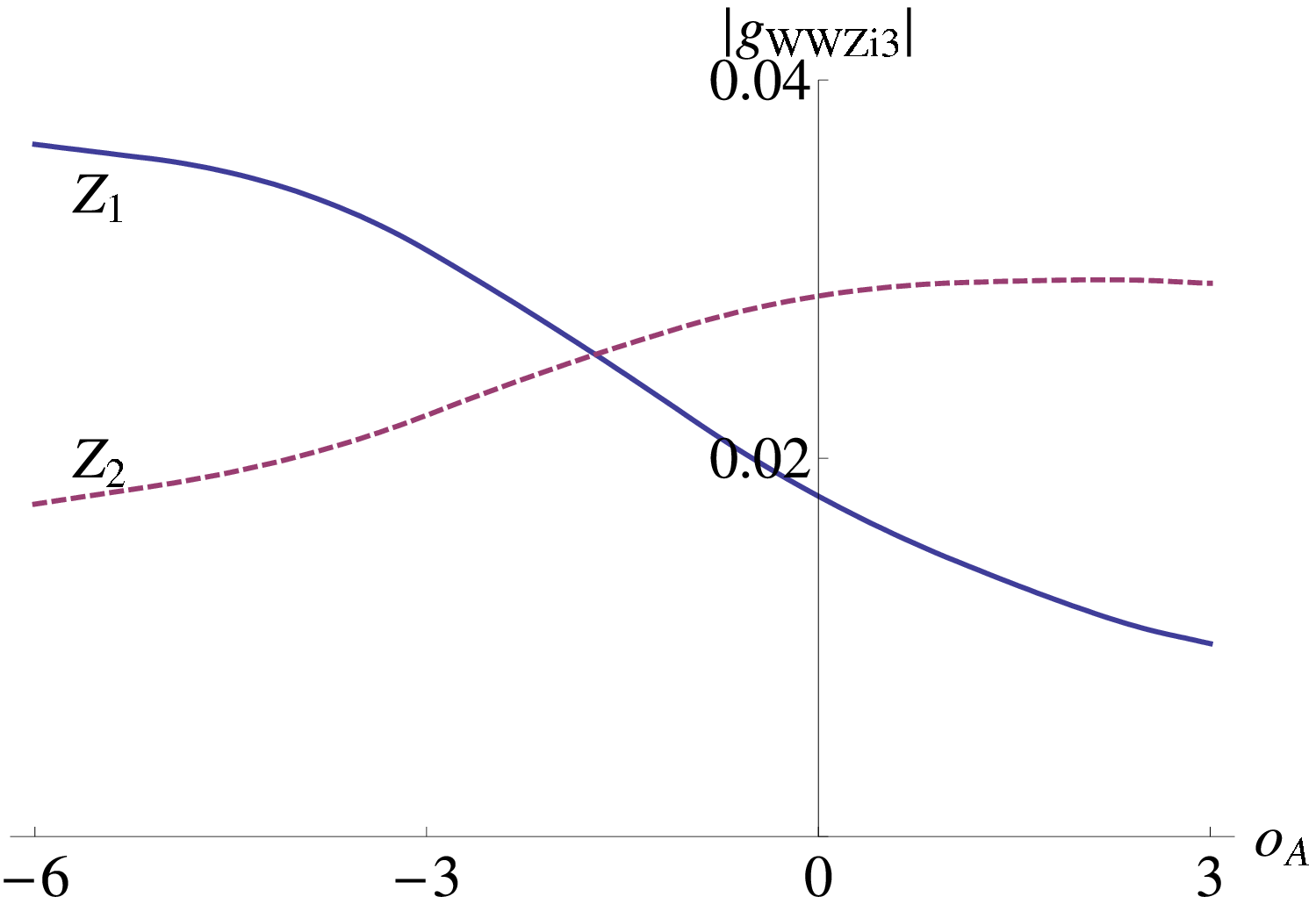}}
  \caption{\label{fig_gWWR3-vertical}Coupling $g_{W W Z_i 3}$ of the two
  lightest neutral resonances $Z_1, Z_2$ to two $W' s$, along the two usual
  curves in parameter space. Other parameters are set in order to produce
  $M_{W_1} = 500 \tmop{GeV}$, as well as the physical values of $\alpha, M_W,
  M_Z$.}
\end{figure}

Figure \ref{fig_gWWR3-vertical} shows, along the two usual lines in parameter
space, the coefficient $g_{W W Z_i 3}$ in the term
\begin{eqnarray}
  - ig_{W W Z_i 3} \partial^{\left[ \mu \right.} Z^{\left. \nu \right]}_i
  W^-_{\mu} W^+_{\nu} . &  &  \label{def-gWWR3}
\end{eqnarray}
Since both light resonances have cubic couplings with two SM gauge fields,
they can both be searched for in $WZ$ final states {\cite{0712.3783}}. In the
case of AdS, we recover that the vector resonances couple predominantly to $W
W$, while the axial ones nearly decouples. Making $o_V$ negative partially
reverses the situation, as expected from Figures \ref{fig_level-crossing} and
\ref{fig_parity}: the second lightest resonance is now more coupled to $W W$
than the first one is.

\subsection{New cubic couplings}\label{sec_permutation}

In the present Section, we detail how different effective warp factors affect
the cubic couplings. In particular, it turns out that $o_A \neq o_V$ allows
for new couplings with interesting phenomenological consequences.

When the two effective warp factors are equal, the TGCs (\ref{TGVs}) satisfy
\begin{eqnarray}
  \kappa_{\gamma} & = & 1, \\
  \kappa_Z & = & g_1^Z . 
\end{eqnarray}
More generally, equality of the two effective warp factors implies that, for
any three spin-1 particles $B, C, D$ the three couplings $g_{B C D 1}, g_{B C
D 2}, g_{B C D 3}$ are equal, according to
\begin{eqnarray}
  g_{B C D 1} &=& g_{B C D 2} = g_{B C D 3} \nonumber\\ & = & - \frac{1}{\sqrt{2}}  \int
  \bignone \frac{\mathd z}{g_5^2} w \left( V_B V_C V_D + V_B A_C A_D + A_B V_C
  A_D + A_B A_C V_D \right),  \label{Adstri}
\end{eqnarray}
where we have already used $w \equiv w_V = w_A$. Applied to photon couplings,
this implies that the photon cannot mediate a transition between two different
particles.

\begin{figure}[hbt]
  \center{\includegraphics[width=7cm]{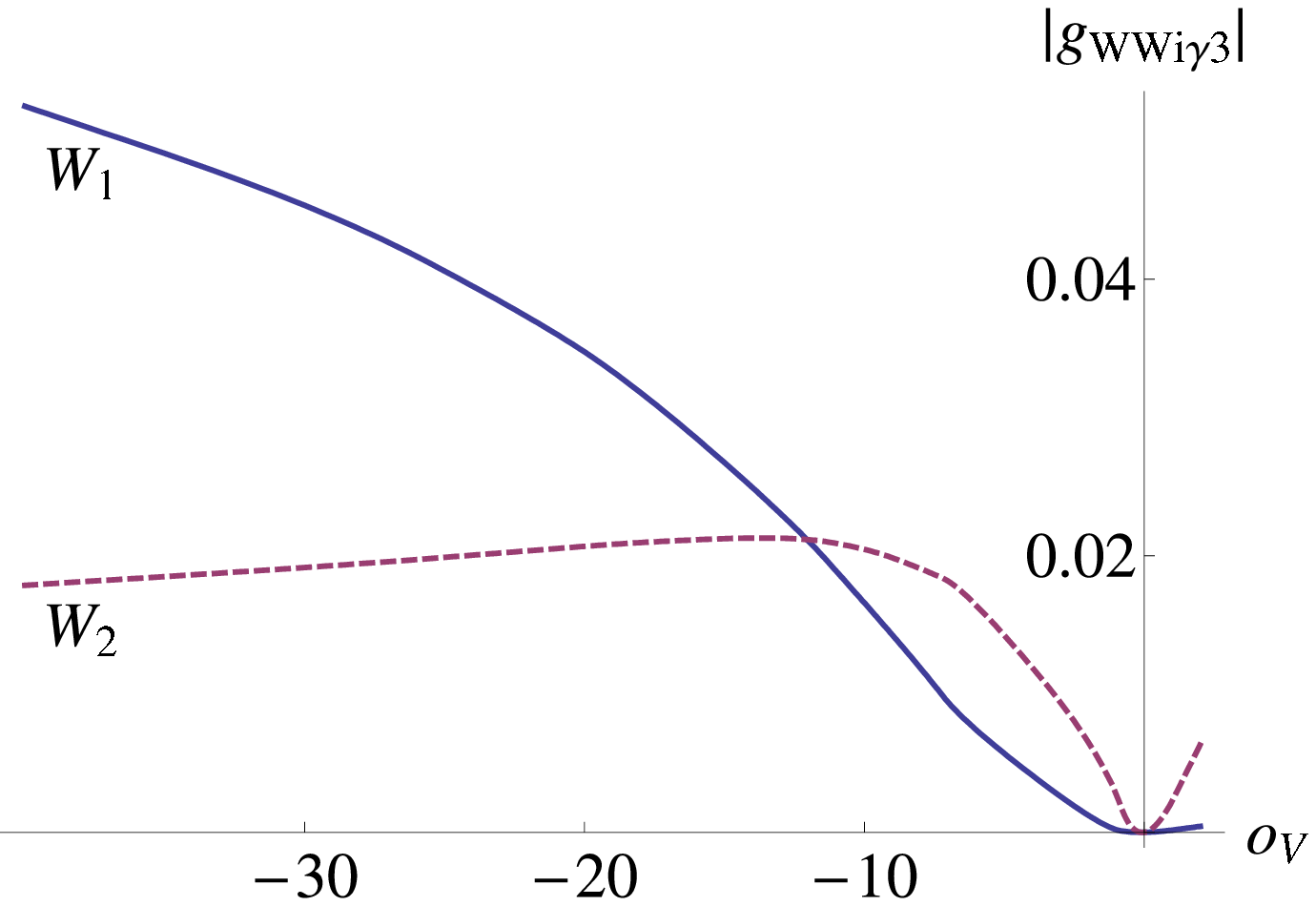}{\hspace{2em}}\includegraphics[width=7cm]{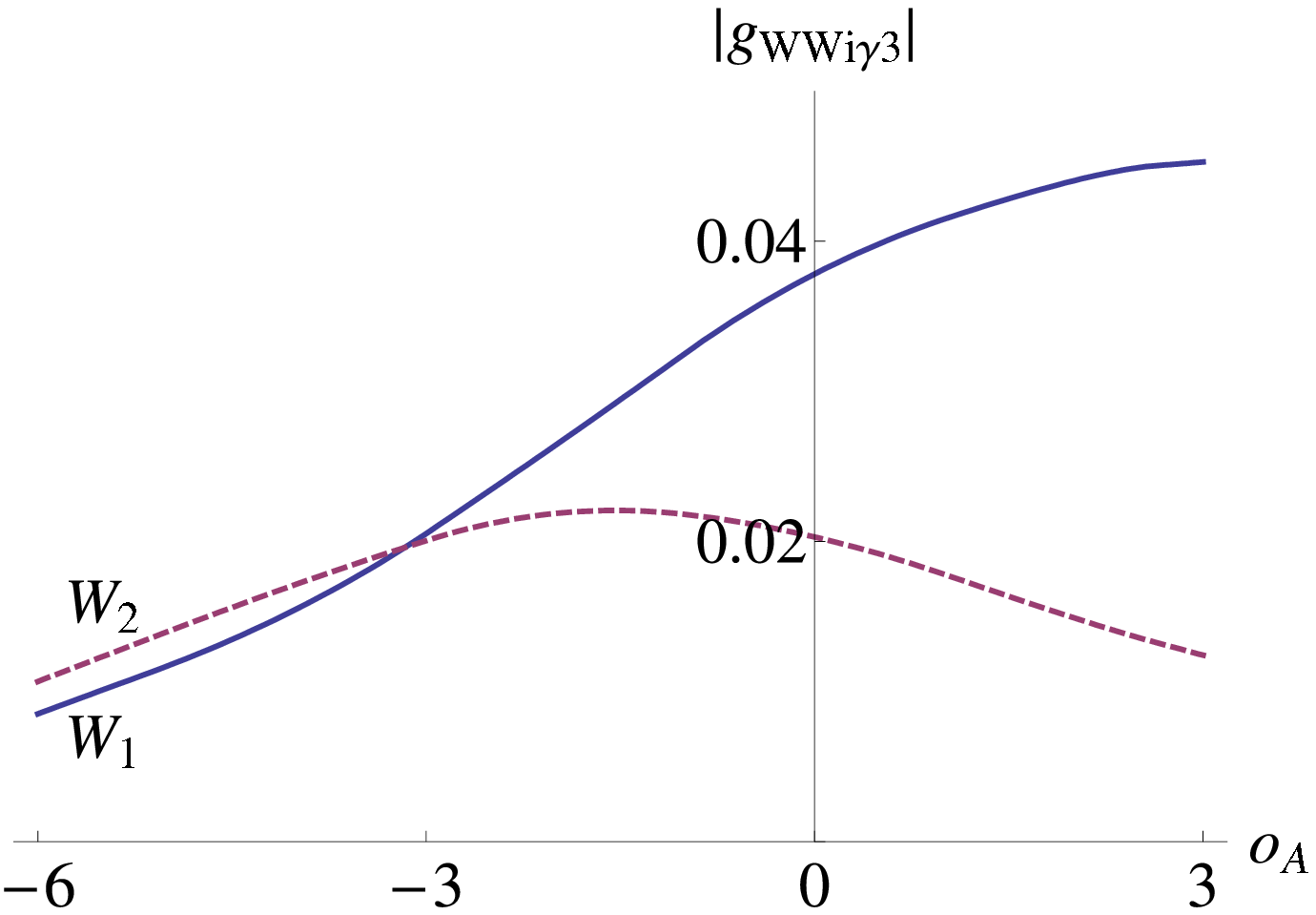}}
  \caption{\label{fig_vertical-couplings}Couplings $g_{W W_i \gamma 3}$ of the
  two lightest charged resonances $W_1, W_2$ to $W \gamma$ along the two usual curves in parameter space. Other
  parameters are set to produce $M_{W_1} = 500 \tmop{GeV}$, as well as the
  physical values of $\alpha, M_W, M_Z$.}
\end{figure}

Deviating from a standard 5D AdS set-up allows for a richer structure. In HTC,
plugging in $V_X ( A_X)$ for the vector (axial) component profile of field $X
\subset \{W^{\pm}, Z, \gamma, W^{\pm}_{1, 2}, Z_{1, 2}^{} \}$, we find
\begin{eqnarray}
  g_{B C D 1} & = & - \frac{1}{\sqrt{2}}  \int  \frac{\mathd z}{g_5^2} (w_V
  (V_B V_C V_D + V_B A_C A_{D_{}}) + w_A (A_B V_C A_D + A_B A_C V_D)), 
  \label{triboson1}\\
  g_{B C D 2} & = & - \frac{1}{\sqrt{2}}  \int  \frac{\mathd z}{g_5^2} (w_V
  (V_C V_D V_B + V_C A_D A_B) + w_A (A_C V_D A_B + A_C A_D V_B)) \bignone, 
  \label{triboson2}\\
  g_{B C D 3} & = & - \frac{1}{\sqrt{2}}  \int  \frac{\mathd z}{g_5^2} (w_V
  (V_D V_B V_C + V_D A_B A_C) + w_A (A_D V_B A_C + A_D A_B V_C)), 
  \label{triboson}
\end{eqnarray}
where $B,$C are charged fields, and $D$ is neutral. In the most general
scenario, where the vector and axial warp factors are different and $B \neq
C^{\ast}, D \neq \gamma$ , the couplings of the three permutations can all be
different
\begin{equation}
  g_{B C D 1} \neq g_{B C D 2} \neq g_{B C D 3} . \label{tri2}
\end{equation}
When $B$ and $C$ are antiparticles of each other, $B$ and $C$ share the same
profile, so the two couplings with the derivative acting on a charged field
are equal
\begin{equation}
  g_{B B D 1} = g_{B B D 2} \neq g_{B B D 3} . \label{tri3}
\end{equation}
When the neutral field is the photon, there is an additional constraint on the
triboson couplings from $U (1)_{\tmop{em}} $ gauge invariance
\begin{equation}
  g_{B C \gamma 1} = g_{B C \gamma 2} = 0 \hspace{1em} \tmop{for} \hspace{1em}
  B \neq C^{\ast} . \label{tri4}
\end{equation}
This result derives from the $\gamma$ wavefunction being flat: the cubic
overlap integrals (\ref{triboson1}-\ref{triboson2}) then reduce to a quadratic
overlap corresponding to the orthogonality relations
\begin{eqnarray}
  \int \bignone \mathd z (w_V V_{\Psi} V_{\Phi} + w_A A_{\Psi} A_{\Phi}) & = &
  0, \hspace{1em} \tmop{for} \hspace{1em} \Psi \neq \Phi .  \label{ortho}
\end{eqnarray}
On the other hand, the third photon coupling (i.e. (\ref{triboson}) with $D =
\gamma$) does not reduce to such an orthogonality relation: it would be
written as (\ref{ortho}), but with the warp factors swapped $w_V
\leftrightarrow w_A$. Therefore, couplings such as  $W^+ W^-_1 \gamma$ vanish
in Higgsless models, but not in HTC. The new coupling is shown in Figure
\ref{fig_vertical-couplings} along the usual lines.

We see two effects at play: first, unequal warp factors $o_A \neq o_V$ allow
for a coupling of the axial resonance to $W \gamma$, and second, the mixing
also allows a coupling of the vector resonance to $W \gamma$. Hence both
resonances can decay to $W \gamma$, as studied in {\cite{0712.3783}}.

\section{Conclusion}

We take a pragmatic approach to the description of dynamical electroweak
symmetry breaking, providing a resonance lagrangian simple enough to be
implemented in Monte-Carlo simulations, yet complex enough to incorporate
phenomena beyond those usually considered. Our description of new strong
interactions does not rely on an explicit model, but rather introduces
phenomenological parameters describing the interactions of new states visible
at the LHC.

Such an effective description usually comes at a price, namely the large
number of unknown parameters to be varied. We deal with this problem by
imposing relations between the constants appearing in the effective
lagrangian: we construct our effective lagrangian using rules from
extra-dimensional model-building in order to impose constraints (we could
equally well have retained a fully 4D formulation by relying on mooses). On
the other hand, we also want to lift some of the 5D constraints that seem too
restrictive. To avoid the alternating spectrum and selection rules usually
predicted by the usual 5D or moose approach, we work with an extension of the
5D framework: Holographic Technicolor (HTC), which starts as a 5D model, but
adds as a new ingredient an effective bulk-breaking term without introducing
new states other than spin-1 resonances.

We do not try to reproduce specific models in the literature, but instead
study the phenomenology of new scenarios that evade some of the usual
constraints on technicolor models. The UV completions of such scenarios are
unknown, but we assume in our effective description that the problem of
oblique corrections is solved, and set the couplings of fermions to $W, Z$
accordingly. We also choose the couplings of SM fermions to resonances to pass
experimental bounds.

We rely on bounds from the trilinear gauge couplings to restrict our three
free parameters, and find that low-mass resonances ($\leqslant 500
\tmop{GeV}$) are allowed. Working in this low-mass assumption, we consider two
interesting curves in the remaining two-parameter space. Two new couplings
that our lagrangian generally includes turn out to be relevant in these
regions, allowing both light resonances to be seen in the $W Z$ channel (as
well as in the $W W$ channel), and to decay to $W \gamma$ (observable at LHC).

Regarding future developments, we point out that our choice of effective
description (drawing on the 5D formalism) allows for an easy inclusion of
additional fields, such as scalars, technipions or the isospin-singlet
techni-omega.

\begin{acknowledgments}
We thank T. Appelquist, R. S. Chivukula and W. Skiba for helpful
comments. JH and AM are supported by DOE grant DE-FG02-92ER-40704; VS
is supported by DE-FG02-91ER40676. JH and VS acknowledge the
hospitality of the Kavli Institute for Theoretical Physics in Santa
Barbara, CA, where part of this work was completed.
\end{acknowledgments}

\appendix

\section{TGCs}\label{app}

In this Appendix, we detail a subtlety in the derivation of TGCs in our
framework as compared to other approaches.

\subsection{Electroweak chiral lagrangian}

The electroweak chiral lagrangian {\cite{Longhitano:1980tm}} describes the
electroweak sector without Higgs at low energies. It is constructed by
coupling the Yang-Mills action of $\tmop{SU} (2)_L \otimes \mathrm{U} \left( 1
\right)_Y$
\begin{eqnarray}
  \mathcal{L}_{\tmop{YM}} & = & - \frac{1}{4}  F^a_{\mu \nu} F^{a, \mu
  \nu} - \frac{1}{4} B_{\mu \nu} B^{\mu \nu},  \label{YM4D}
\end{eqnarray}
to a chiral lagrangian for the three Goldstone bosons (GBs) of the breaking
$\tmop{SU} (2) \times \tmop{SU} (2) \rightarrow \tmop{SU} (2)$ {\emdash}this
is the minimal custodial sector that feeds 3 GBs without introducing
technipions. This GB chiral lagrangian, ordered by the number of derivatives,
starts with the operator{\footnote{In the present framework, $T$
    vanishes, therefore we do not write the corresponding operator.}}
\begin{eqnarray}
  \mathcal{L}_{D^2} & = & \frac{f^2}{4}  \left\langle D_{\mu} U^{\dag} D^{\mu}
  U \right\rangle,  \label{LD2}
\end{eqnarray}
where $\left\langle \right\rangle$ means $\tmop{SU}(2)$ trace and the covariant derivative applied to the GB unitary matrix $U$ reads
\begin{eqnarray}
  D_{\mu} U & = & \partial_{\mu} U - igW_{\mu} U + ig' B_{\mu} U
  \frac{\tau^3}{2} .  \label{DmuU}
\end{eqnarray}
The covariant derivative couples the weak $\tmop{SU} (2)_L \otimes \mathrm{U}
\left( 1 \right)_Y$ gauge fields to the GBs. This can be checked by going to
unitary gauge, i.e. $U = \mathbbm{1}$, which yields the appropriate masses for
the $W^{\pm}$ and $Z^0$.

At the next order in the derivative expansion, there are extra quadratic
($\alpha_1$), cubic ($\alpha_2, \alpha_3$) and quartic ($\alpha_4, \alpha_5$)
couplings among the $\gamma, W, Z$
\begin{eqnarray}
  \mathcal{L}_{D^4} & = & \alpha_1 gg' B_{\mu \nu}  \left\langle U
  \frac{\tau^3}{2} U^{\dag} W^{\mu \nu} \right\rangle \nonumber\\
  & + & i \left\langle \left( \alpha_2 gW_{\mu \nu} + \alpha_3 g' B_{\mu \nu}
  \frac{\tau^3}{2} \right)  \left[ UD^{\mu} U^{\dag}, UD^{\nu} U^{\dag}
  \right] \right\rangle \nonumber\\
  & + & \alpha_4  \left\langle D_{\mu} UD_{\nu} U^{\dag} \right\rangle^2 +
  \alpha_5  \left\langle D_{\mu} UD^{\mu} U^{\dag} \right\rangle^2 . 
  \label{Op4}
\end{eqnarray}
Assuming that the underlying strong dynamics respect parity symmetry results
in $\alpha_2 = \alpha_3$.{\footnote{The Longhitano couplings $\alpha_i$ are related to the
Gasser-Leutwyler $L_i$ couplings (which assume parity invariance) by $\alpha_1
= L_{10}$, $\alpha_2 = \alpha_3 = - L_9 / 2$. The $\alpha_i$ coefficients with
$i > 5$ introduce isospin breaking at tree level, and
will therefore not be generated in our framework.}}

The $\alpha_1$ term introduces the mixing of $W_{\mu}^3$ and $B_{\mu}$, even
though the quadratic lagrangian may have been diagonalized at the previous
order; $\alpha_1$ is thus an oblique correction. One may choose to diagonalize
 the full quadratic lagrangian at this order (i.e. including $\alpha_1$) in
order to work in the mass basis, thereby  shifting the deviation from the SM
into fermion couplings {\cite{hep-ph/9907294}}.

In the framework of HTC, $\alpha_1$ is computed as follows
\begin{eqnarray}
  \alpha_1 & = & - \frac{1}{2}  \int_{l_0}^{l_1} \frac{\mathd z}{g_5^5} 
  \left( w_V (z) - w_A (z) \left( \frac{\int_z^{l_1} \mathd z' / w_A \left( z'
  \right) \bignone}{\int_{l_0}^{l_1} \mathd z'' / w_A \left( z'' \right)
  \bignone} \right)^2 \right) \bignone, 
\end{eqnarray}
which is the expression used to determine the shape of line B in Figure
\ref{fig_alpha1_0}.

The cubic terms in $\alpha_1$ as well as $\alpha_2, \alpha_3$ in the
lagrangian (\ref{Op4}) produce deviations from the SM in the TGCs
(\ref{TGVs}), given by {\cite{Holdom:1990xq}} (assuming $\alpha_2 = \alpha_3$)
\begin{eqnarray}
  \kappa_{\gamma} - 1 & = & \frac{e^2}{s^2}  \left( - \alpha_1 + 2 \alpha_3
  \right),  \label{kappagamma}\\
  g_1^Z - 1 & = & \frac{e^2}{c^2}  \left( \frac{\alpha_1}{c^2 - s^2} +
  \frac{\alpha_3}{s^2} \right),  \label{g1Z}\\
  \kappa_Z - 1 & = & e^2  \left( \frac{2 \alpha_1}{c^2 - s^2} + \frac{c^2 -
  s^2}{c^2 s^2} \alpha_3 \right) .  \label{kappaZ}
\end{eqnarray}
This derivation uses the standard relation
\begin{eqnarray}
  G_F & = & \frac{g^2}{4 \sqrt{2} M_W^2} \hspace{1em} = \hspace{1em}
  \frac{1}{\sqrt{2} f^2},  \label{GFf}
\end{eqnarray}
which is valid for the electroweak chiral lagrangian, but not true in general.
Indeed, relation (\ref{GFf}) assumes that $g$ appearing in the covariant derivative
(\ref{DmuU}) is strictly equal to the fermion-$W$ coupling, something which
does not hold in our framework, nor in standard 5D models
{\cite{hep-ph/0510275}}.

\subsection{TGCs in Holographic Technicolor}\label{sec_subtlety}

We chose to {\tmem{define}} the interactions of the fermions by assigning them
couplings to the physical mass eigenstates that satisfy the SM relations.
Therefore, the $S$ and $T$ parameters will vanish. On the other hand, we model
the interactions of spin-1 fields, imposing experimental values for the
parameters $\alpha, M_Z, M_W$, and predicting the TGCs. As a consequence,
relations between fermion couplings (which are set by hand) and quantities
involving $\gamma$, $W, Z$ (which are predicted from a 5D lagrangian) are
modified. For example
\begin{eqnarray}
  M_W & \neq & \frac{gf}{2},  \label{MWgf}
\end{eqnarray}
does not hold at tree level if $g$ means the fermion-$W$ coupling --this
coupling is set by hand outside any 5D modelling, whereas $M_W$ and $f$ are an
output of the 5D model. In other words, we have in general
\begin{eqnarray}
  f & \neq & 246 \tmop{GeV} . 
\end{eqnarray}

\begin{figure}[hbt]
  \includegraphics[width=17cm]{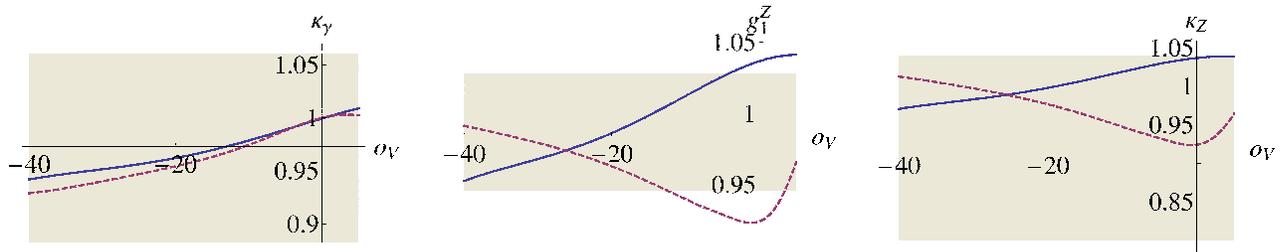}
  \caption{\label{fig_relation-Holdom}TGCs along line A, for $M_{W_1} = 500
  \tmop{GeV}$, with all other parameters at their physical value. The
  horizontal axis depicts the central value measured at LEP, while the shaded
  bands depict the $2 \sigma$ errors. The continuous lines depict the TGCs as
  a function of $o_V$, for $o_A = 0$ and $M_{W_1} = 500 \tmop{GeV}$. Dashed
  lines represent values computed from blindly using
  (\ref{kappagamma}-\ref{g1Z}). The discrepancy in the $\kappa_{\gamma}$ plot
  comes from having used the $l_0 \rightarrow 0$ approximation for the dashed
  line.}
\end{figure}

As a consequence, we find that, while relation (\ref{kappagamma}) holds in our
framework, the same is not true of (\ref{g1Z}-\ref{kappaZ}). This is because
these latter two definitions depend on the value of the Weinberg angle,
defined as
\begin{eqnarray}
  s^2 c^2 & = & \frac{\pi \alpha}{\sqrt{2} G_F M_Z^2} . 
  \label{Weinberg-angle}
\end{eqnarray}
The extraction of the TGCs thus involves $G_F$ and not $f$ whereas Eqs.
(\ref{kappagamma}-\ref{kappaZ}) assumed a relation between the two,
Eq. (\ref{GFf}).
The definition of the Weinberg
angle (\ref{Weinberg-angle}) feeds into the extraction of the TGCs from the
lagrangian (\ref{TGVs}). Figure \ref{fig_relation-Holdom} compares the exact
prediction of the TGCs with the one that would be derived from
(\ref{kappagamma}-\ref{kappaZ}).

To understand the discrepancy, we must go back to the definitions of
$G_F$
\begin{eqnarray}
  G_F & = & \frac{1}{4 \sqrt{2}}  \left( \frac{g^2}{M_W^2} + \sum_{n = 1}
  \frac{g_n^2}{M_{W_n}^2} \right),  \label{exactGF}
\end{eqnarray}
where $g, g_n$ are the fermion-spin-1 couplings,
and  $f$ {\cite{hep-ph/0702005}}
\begin{eqnarray}
  \frac{1}{f^2} & = & 2 \sum_{n = 0} \frac{A_{W_n} (l_0)^2}{M_{W_n}^2}. 
  \label{relf}
\end{eqnarray}
There is no connection between $f$ and $G_F$ unless the fermions are modeled in 5D in
a particular way. If, for example, the fermions were placed on the UV brane,
then there would be a relation between the $g_n$'s of (\ref{exactGF}) and the
wavefunctions  $A_n (l_0)$ of the resonances on the UV brane.

\end{document}